\documentstyle[epsfig,aps,twocolumn]{revtex}

\newcommand{\pol}{\hat{\bf e}}
\newcommand{\rv}{{\bf r}}
\newcommand{\Ev}{{\bf E}}
\newcommand{\Dv}{{\bf D}}
\newcommand{\Pv}{{\bf P}}

\newcommand{\dv}{{\bf d}}

\newcommand{\Dc}{{\cal D}}

\newcommand{\Hc}{{\cal H}}
\newcommand{\Fc}{{\cal F}}

\newcommand{\kv}{{\bf k}}

\newcommand{\eo}{\epsilon_0}
\newcommand{\beq}{\begin{equation}}
\newcommand{\eeq}{\end{equation}}
\newcommand{\bea}{\begin{eqnarray}}
\newcommand{\eea}{\end{eqnarray}}
\newcommand{\<}{\langle}
\renewcommand{\>}{\rangle}
\renewcommand{\(}{\left(}
\renewcommand{\)}{\right)}
\renewcommand{\[}{\left[}
\renewcommand{\]}{\right]}

\newcommand{\up}{\uparrow}
\newcommand{\down}{\downarrow}
\newcommand{\commentout}[1]{{}}

\begin{document}

\draft
\preprint{}
\title{Optical response of superfluid state in dilute 
atomic Fermi-Dirac gases} 
\author{J. Ruostekoski} 
\address{Abteilung f\"ur Quantenphysik,
Universit\"at Ulm, D-89069 Ulm, Germany}
\date{\today}
\maketitle
\begin{abstract}
We theoretically study the propagation of light in a 
Fermi-Dirac gas in the presence of a superfluid state.
BCS pairing between atoms in different hyperfine levels
may significantly increase the optical linewidth and line shift of a
quantum degenerate Fermi-Dirac gas
and introduce a local-field correction that, under certain conditions,
dramatically dominates over the Lorentz-Lorenz shift.
These optical properties could possibly unambiguously sign the presence 
of the superfluid state and determine the value of the BCS order 
parameter.

\end{abstract}

After the first observations of atomic Bose-Einstein condensates 
\cite{becexp} there has been an increasing interest in
the studies of Fermi-Dirac (FD) gases
\cite{JAV95b,STO96,HOU97,MAR98,BAR96,BRU98b,JAV99,RUO99c}. 
One especially fascinating property of FD gases is
that with effectively attractive interaction between different particles the
ground state of the system may become unstable with respect to formation of
bound pairs of quasi-particles or the Cooper pairs \cite{LIF80,FET71}. 
This effect is analogous to the BCS transition in superconductors. The
particles near the Fermi surface having opposite momenta and different internal
quantum number tend to appear in pairs. This leads, e.g., to a finite energy
gap in the excitation spectrum of the system and to a nonvanishing 
anomalous expectation value $\< \psi_\uparrow (\rv)
\psi_\downarrow (\rv) \>$. Here the two internal sublevels are referred to
as $\up$ and $\down$.

In this paper we study the optical response of a superfluid state in 
a zero-temperature FD gas at low atom densities for low-intensity 
light. We show
that the presence of the coherent pairing between atoms in different
hyperfine levels may dramatically {\it enhance} optical interactions 
and the scattering of light in FD gas.

One particularly promising candidate to
undergo the BCS transition and to become a superfluid is spin-polarized
atomic $^6$Li. Atoms in two different internal levels can interact 
via
$s$-wave scattering and the $^6$Li atom has an anomalously large and negative
$s$-wave scattering length $a\simeq-2160a_0$, where $a_0$ is 
the Bohr radius.
The nuclear spin states $m_i=1$ and 0 of $^6$Li have been 
predicted to undergo a superfluid transition at $10^{-8}$ K with 
the density of $10^{12}$ cm$^{-3}$ \cite{STO96,HOU97}.  

We study the propagation of light by introducing
the dipole approximation for atoms and the corresponding 
Hamiltonian in the {\it length} gauge obtained in the Power-Zienau-Woolley 
transformation \cite{COH89}.
FD gas is assumed to occupy two different hyperfine levels
$|g,\up\>$ and $|g,\down\>$ of the same atom. For simplicity,
we consider here a situation where there are only two electronically excited
levels $|e,\up\>$ and $|e,\down\>$. All the dipole matrix elements for optical
transitions between the levels are assumed to vanish, except 
$\dv_{g\up e\up}$ and $\dv_{g\down e\down}$, where $\dv_{g\up e\up}$
denotes the dipole matrix element for the transition $|e,\up\>\rightarrow 
|g,\up\>$. For instance, the electric dipole moment between levels 
$\up$ and $\down$ vanish, if the levels
refer to different nuclear spin states that are decoupled from the levels
involved in optical transitions by an external magnetic field.

In the absence of the driving light field atoms in the electronic ground
state are described in second quantization
by the Hamiltonian density $\Hc_1$ \cite{LIF80,FET71}:
\beq
\Hc_1=\sum_\nu \psi^\dagger_{g\nu}(H^{g\nu}_{\rm c.m.}-\mu_{g\nu})
\psi_{g\nu}+ \hbar u_g \psi^\dagger_{g\up}\psi^\dagger_{g\down}
\psi_{g\down}\psi_{g\up}\,,
\label{ground}
\eeq
where $\psi_{g\nu}(\rv)$ is the atom field operator for level $|g,\nu\>$
in the Heisenberg picture, $\mu_{g\nu}$ is the corresponding chemical
potential, and $H^{g\nu}_{\rm c.m.}$ denotes the center-of-mass (c.m.)
Hamiltonian.
We have approximated the finite-range interparticle potential by
a contact interaction with the strenght given by 
$u_g=4\pi a_g\hbar/m$, where $a_g$ is the $s$-wave scattering length and
$m$ is the mass of the atom. The atoms in different hyperfine 
levels can interact via $s$-wave scattering. On the other hand,
there only is a very weak $p$-wave scattering between two atoms in the same
hyperfine state which is ignored in Eq.~{(\ref{ground})}.

The driving light field introduces additional terms for the system 
Hamiltonian.
In the length gauge the basic dynamical degree of freedom for the light 
field is the electric displacement $\Dv(\rv)$ that interacts with the 
atomic polarization $\Pv(\rv)$
\beq
\Hc_2=-{1\over\eo}\Pv(\rv)\cdot\Dv(\rv)\,.
\label{h2}
\eeq
In the present case the positive frequency component 
of the polarization is given by
\bea
\Pv^+(\rv)&=& \dv_{g\up e\up}\psi^\dagger_{g\up}(\rv)\psi_{ e\up}(\rv) +
 \dv_{g\down e\down}\psi^\dagger_{g\down}(\rv)\psi_{ e\down}(\rv) 
\nonumber\\
&\equiv&
\Pv^+_{\up}(\rv)+\Pv^+_{\down}(\rv)\,.
\label{pol}
\eea

The polarization self-energy was shown in Ref. \cite{RUO97c}
to be inconsequential for dipole atoms.
We assume that to leading order all remaining interactions 
between the atoms and the light field,
that cannot be accounted for when the atoms are modeled 
as point dipoles,
are governed by the following interactions:
\bea
\Hc_3 &=& \sum_\nu \psi^\dagger_{e\nu}(H^{e\nu}_{\rm c.m.}+
\hbar\omega_0-\mu_{e\nu})\psi_{e\nu}+
\hbar u_e \psi^\dagger_{e\up}\psi^\dagger_{e\down}
\psi_{e\down}\psi_{e\up}\nonumber\\
&&+\sum_{\nu,\sigma}\hbar u_{g\nu e\sigma}\psi^\dagger_{g\nu}
\psi^\dagger_{e\sigma}\psi_{e\sigma}\psi_{g\nu}\,.
\label{h3}
\eea
Here $u_{g\nu e\sigma}=4\pi\hbar a_{g\nu e\sigma}/m$ and 
$u_{e}=4\pi\hbar a_{e}/m$ describe the two-body $s$-wave
scattering between the atoms. For simplicity,
the frequency of the optical transition $\omega_0$ is assumed to
be independent of the hyperfine level. For typical values of the
optical linewidth the c.m.\ motion for the excited atoms may be omitted
\cite{JAV95b}. 

The positive frequency component of the
electric field $\Ev^+$ may be obtained by solving the Heisenberg
equations of motion~\cite{RUO97a} 
\begin{mathletters}
\bea
\eo{\bf E}^+({\bf r})& =& {\bf D}^+_F({\bf r}) +
\int d^3r'\,
{\sf G}({\bf r}-{\bf r'})\,{\bf P}^+({\bf r}')\,,
\label{eq:MonoD}\\
{\sf G}_{ij}({\bf r})& =& 
\left[ {\partial\over\partial r_i}{\partial\over\partial r_j} -
\delta_{ij} {\bbox \nabla}^2\right] {e^{ikr}\over4\pi r}
-\delta_{ij}\delta({\bf r})
\,.
\label{eq:GDF}
\eea
\label{eq:atomlight}
\end{mathletters}
Here ${\bf D}^+_F$ is the positive frequency component of the driving 
electric displacement with the frequency $\Omega$, and
$k = {\Omega / c}$. The monochromatic dipole radiation
kernel ${\sf G}(\rv)$ is precisely the classical expression of the dipolar
field, including the delta function at the origin \cite{JAC75}.

In the limit of low light intensity we have derived from the Heisenberg
equations of motion a hierarchy of
equations for correlation functions involving atomic
polarization and atom density \cite{RUO97a,RUO97c}. In the case of the 
present system we may proceed similarly. As far as the optical response 
is concerned it is again assumed that we can concentrate on the dynamics of
internal degrees of freedom for the atoms and the light. Hence, in the
equation of motion for the atomic polarization 
the kinetic energy of the atoms is neglected. 

Light has to be present in order to produce population in the 
electronically excited levels. In the present paper we consider
the limit of low intensity of the driving light.
This is done by retaining only those products of operators that 
involve at most one excited state field operator or
the driving electric displacement \cite{RUO97a}. Then, e.g., 
the term proportional to $u_e$ in Eq.~{(\ref{h3})} has no 
contribution to the equation of motion for $\Pv^+(\rv)$.

To simplify further we assume that $\dv_{g\up e\up}=\dv_{g\down e\down}$.
It is useful to introduce the following projection operator 
\beq
{\sf P}\equiv \dv_{g\up e\up}\dv_{g\up e\up}/|\dv_{g\up e\up}|^2\,,
\eeq
that projects the internal degrees of freedom 
onto the subspace defined by the four hyperfine levels in consideration.  
For the expectation values we use the notation $\Pv_{1\nu}\equiv
\<\Pv^+_{\nu}\>$, for $\nu$ denoting the hyperfine state. 
The steady-state solution of $\Pv_{1\nu}$ is given by
\bea
\lefteqn{
\Pv_{1\nu}(\rv_1)=\alpha\rho_{\nu }{\sf P}\cdot\Dv^+_F(\rv_1)+
\sum_\sigma\Fc_{\sigma\nu}\Pv_2(\rv_1\sigma ;\rv_1\nu )}
\nonumber\\ &&+ \mbox{}
\alpha\sum_\sigma
\int d^3 r_2 \, {\sf P}\cdot{\sf G'}(\rv_1-\rv_2)
\Pv_2(\rv_1\nu ;\rv_2\sigma )\,.
\label{p1}
\eea
Here $\alpha=-\Dc^2/[\hbar\eo(\delta+i\gamma)]$ is the 
polarizability of a single atom, 
${\cal D}$ is the reduced dipole matrix element,
$\rho_\nu$ denotes the atom density in level $\nu$,
$\gamma=\Dc^2 k^3/(6\pi\hbar\eo)$ is the spontaneous
linewidth, and $\delta$ the atom-light detuning. We have also defined
\bea
&&\Pv_2(\rv_1\nu;\rv_2\sigma) \equiv \< \psi^\dagger_{g\nu}(\rv_1)
\Pv^+_{\sigma}(\rv_2)\psi_{g\nu}(\rv_1)\>,\\
&&\Fc_{\sigma\nu} \equiv {1\over \delta+i\gamma}[u_{g\sigma e\nu}
-(1-\delta_{\sigma\nu})u_g]\,.
\eea
The normally ordered expectation value $\Pv_2(\rv_1\nu;\rv_2\sigma)$ 
describes correlations between an atomic dipole at $\rv_2$ in hyperfine 
level $\sigma$ and a ground state atom at $\rv_1$ in hyperfine 
level $\nu$. The tensor $\Fc_{\sigma\nu}$ generates the 
collisionally-induced level shifts.

Due to the hard-core interatomic potential we
remove the contact dipole-dipole interactions between different
atoms. In Eq.~{(\ref{p1})} this is done by
introducing the propagator $
{\sf G}_{ij}'(\rv)={\sf G}_{ij}(\rv)+\delta_{ij}\delta(\rv)/3
$.
The purpose of this definition is to yield a vanishing integral 
for ${\sf G}'(\rv)$ 
over an infinitesimal volume enclosing the origin \cite{RUO97c}.

So far, we have obtained a steady-state solution for the atomic 
polarization {(\ref{p1})} that acts as a source for the secondary
radiation in Eq.~{(\ref{eq:MonoD})}. Equation {(\ref{p1})} involves
unknown correlation function $\Pv_2$. Basically, we could continue
the derivation and obtain the equations of motion for $\Pv_2$ and
for the higher order correlation functions. This would eventually result 
in an infinite hierarchy of equations analogous to the equations in Ref.
\cite{RUO97a}. However, even in the case of a simple level structure 
and in the absence of the $s$-wave interactions the solution for the whole 
system by stochastic simulations is demanding on computer time 
\cite{JAV99}. In the studies of the refractive index of a quantum 
degenerate Bose-Einstein gas Morice {\it et al.} \cite{MOR95} 
derived a density expansion in terms of the number of atoms 
repeatedly exchanging a photon by introducing certain approximations
in the ground state atom correlations.
Although the lowest order density correction for the electric 
susceptibility of a zero-temperature FD gas may be obtained
analytically \cite{RUO99c}, in the presence of nontrivial 
statistical position correlations a rigorous density expansion is
in most cases a very challenging task. 
In this paper we consider low atom densities (in terms of
$\rho/k^3$) and approximate 
Eq.~{(\ref{p1})} by the decoupling that corresponds to the lowest order 
correction in Ref. \cite{MOR95}
\beq
\Pv_2(\rv_1\nu;\rv_2\sigma)\simeq\rho_2(\rv_1\nu,\rv_2\sigma)
\Pv_{1\sigma}(\rv_2)/\rho_\sigma\,,
\label{p2}
\eeq
where the ground state pair correlation function $\rho_2$ is defined by
\beq
\rho_2(\rv_1\nu,\rv_2\sigma)=\< \psi^\dagger_{g\nu}(\rv_1)
\psi^\dagger_{g\sigma}(\rv_2)\psi_{g\sigma}(\rv_2)
\psi_{g\nu}(\rv_1) \>\,.
\label{pair}
\eeq
It is important to point out that
the predictions of the expansion by Morice {\it et al.} \cite{MOR95} 
were tested for a zero-temperature FD gas in one dimension \cite{JAV99}. 
The agreement with the exact solution obtained by the numerical 
simulations was found to be semi-quantitative and in the low-density 
limit excellent.

Before the light is switched on, the system is described by the
Hamiltonian density $\Hc=\Hc_1$ [Eq.~{(\ref{ground})}]. 
The assumption that the driving light only weakly disturbs the 
system allows us to evaluate the pair correlation functions for
the ground state atoms [Eq.~{(\ref{pair})}]
from $\Hc_1$ even in the presence of the driving light. 
We assume a homogeneous sample and introduce a
plane wave basis for the field operators: $\psi_{g\nu}(\rv)=V^{-1/2}
\sum_{\kv}b_{\kv\nu}\exp(i\kv\cdot \rv)$. 
The Hamiltonian {(\ref{ground})} is diagonalized by the standard canonical 
transformation to the Bogoliubov quasi-particles \cite{LIF80,FET71}  
\begin{mathletters}
\bea
\alpha_\kv &=& u_\kv b_{\kv\down}-v_\kv b_{-\kv\up}^\dagger ,\\
\beta_{-\kv} &=& u_\kv b_{-\kv\up}+v_\kv b_{\kv\down}^\dagger\,,
\eea
\label{bogo}
\end{mathletters}
where $u_\kv$ and $v_\kv$ are real, depend only only on $|\kv|$, and satisfy
$u_\kv^2+v_\kv^2=1$. 
The requirement that $\Hc_1$ in Eq.~{(\ref{ground})} is diagonal in the
quasi-particle representation sets an additional constraint and we obtain
\beq
u_\kv^2 = {1\over 2}\(1+{\xi_\kv\over E_\kv}\), \quad
v_\kv^2 = {1\over 2}\(1-{\xi_\kv\over E_\kv}\) \,,
\label{bogo2}
\eeq
where $E_\kv=\sqrt{\Delta^2+\xi_\kv^2}$, $\xi_\kv =\epsilon_{\kv}-
\bar{\mu} +\hbar u_g (\rho_\up+\rho_\down)/2$, and the energy gap 
$\Delta=-\hbar u_g V^{-1} \sum_\kv u_\kv v_\kv(1-\bar{n}_{\alpha\kv}
-\bar{n}_{\beta\kv})$. In equilibrium, the quasi-particle occupation numbers 
$\bar{n}_{\alpha\kv}\equiv \< \alpha_\kv^\dagger \alpha_\kv \>$ and
$\bar{n}_{\beta\kv}\equiv \< \beta_\kv^\dagger \beta_\kv \>$ satisfy
FD statistics with $\bar{n}_{\alpha\kv}=\bar{n}_{\beta\kv}=
(e^{E_\kv/k_BT}+1)^{-1}$.
The dispersion relation for free particles
is given by $\epsilon_{\kv}=\hbar^2 k^2/(2m)$ and
the average of the chemical potentials is $\bar{\mu}=(\mu_\up+\mu_\down)/2$.
For simplicity, from now on we assume $\mu_\up=\mu_\down$.

In the superfluid phase transition the atoms in the different hyperfine
levels $\up$ and $\down$ form a quasi-particle condensate that results
in a nonvanishing anomalous correlation 
$\<\psi_\up(\rv_1)\psi_\down(\rv_2)\>$.
The effect of this macroscopic two-particle coherence on 
the pair correlation 
function (\ref{pair}) can be clearly seen by considering the ground
state of $\Hc_1$ [Eq.~{(\ref{ground})}] that is the vacuum of 
the Bogoliubov quasi-particles [Eq.~{(\ref{bogo})}]. Then
(for $\nu\neq\sigma$)
\begin{mathletters}
\bea
\rho_2(\rv_1\nu,\rv_2\sigma)&=& \rho_\nu\rho_\sigma+
|\< \psi_{g\nu}(\rv_1)\psi_{ g\sigma}(\rv_2) \>|^2,\\
\rho_2(\rv_1\nu,\rv_2\nu)&=& \rho_\nu^2-
|\< \psi^\dagger_{g\nu}(\rv_1)\psi_{ g\nu}(\rv_2) \>|^2\,.
\eea
\label{gaus}
\end{mathletters}

The optical response may now be evaluated by eliminating $\Dv_F^+$ and
$\Pv_2$ from Eqs.~{(\ref{eq:MonoD})}, {(\ref{p1})}, and {(\ref{p2})}.
Because we are dealing with a linear theory, the electric field and the
polarization are related by the susceptibility as 
$\Pv^+=\eo\chi \Ev^+$.
We consider a situation where FD gas fills the half-infite
space $z>0$. We assume that the constant atom densities for the hyperfine 
states are equal $\rho_\up=\rho_\down\equiv\rho$. For simplicity,
it is also assumed that the $s$-wave scattering length $a_{g\nu e\sigma}$
is independent of hyperfine states $\nu$ and $\sigma$ resulting in
$\Fc_{\up\up}=\Fc_{\down\down}$ and $\Fc_{\up\down}=
\Fc_{\down\up}$.
The incoming free field is written $\Dv_F(\rv)=
D_F \,\pol \exp{(ikz)}$, and we assume that the polarization $\pol$
is parallel to the electric dipole moments $\dv_{g\up e\up}$ and
$\dv_{g\down e\down}$.
With the ansatz $\Pv_{1\up}(\rv)=\Pv_{1\down}(\rv)=P\,
\pol\exp{(ik'z)}$ for Im$(k')>0$, by using Eq.~{(\ref{gaus})},
and by ignoring the effects of the surface of the atomic gas 
\cite{JAV99}, we obtain a spatially constant susceptibility for
the sample as
\begin{equation}
\chi={{k'}^2\over k^2} -1= {2\alpha\rho\over1-2\alpha\rho/3+
\Sigma_1+\Sigma_2}\,,
\label{sus}
\end{equation}
with
\bea
\Sigma_1 &=&-{\alpha\over\rho}\int d^3r\,e^{-ikz}
{\sf G'}(\rv) \[ |\< \psi_{g\up}(\rv)\psi_{ g\down}(0) \>|^2\right.
\nonumber\\
&&\left.-|\< \psi^\dagger_{g\down}(\rv)\psi_{ g\down}(0) \>|^2\],
\label{c}
\\
\Sigma_2 &=& -{1\over\rho}\sum_{\nu,\sigma} \Fc_{\nu\sigma}
\rho_2(\rv\nu,\rv\sigma)\,.
\label{colshift}
\eea
Here we have used the obvious relations
$\rho_2(\rv_1\up,\rv_2\down)=\rho_2(\rv_1\down,\rv_2\up)$ and
$\rho_2(\rv_1\up,\rv_2\up)=\rho_2(\rv_1\down,\rv_2\down)$. 

In an uncorrelated atomic sample the atomic positions are
statistically independent and the pair correlation function
(\ref{pair}) satisfies $\rho_2(\rv\nu,\rv'\sigma)=\rho_\nu
\rho_\sigma$. In this case, and in the absence of the $s$-wave 
scattering,
we would obtain Eq.~{(\ref{sus})} with $\Sigma_1=\Sigma_2=0$. 
This is the standard column density
result stating that susceptibility equals polarizability of an
atom times atom density. Equation {(\ref{sus})} also involves 
the Lorentz-Lorenz local-field correction in the denominator.

The quantum statistical corrections to the column density result
are introduced by $\Sigma_1$. It describes
the modifications of the optical interactions between neighboring
atoms due to the position correlations.
The second term in Eq.~{(\ref{c})} represents the quantum statistical 
contribution to the scattering process in which a photon 
emitted by an atom in hyperfine level $\nu$ at position $\rv$
is reabsorbed by another atom in hyperfine level $\nu$ and 
located at the origin. 
According to FD statistics two fermions with the same quantum 
numbers repel each other and FD statistics forces a regular 
spacing between the atoms.
The optical interactions are dominantly generated at small 
interatomic distances and the
corrections to the susceptibility due to the second term 
in Eq.~{(\ref{c})} correspond to {\it inhibited}
light scattering. In the absence of superfluid state
FD gas exhibits a dramatic line narrowing
\cite{JAV99,RUO99c}.

The first term in Eq.~{(\ref{c})}
represents the quantum statistical corrections to the reabsorption
process between atoms in different hyperfine levels due to the 
two-particle coherence. This term is nonvanishing
only in the presence of a superfluid state. Because the atom pairs in
Eq.~{(\ref{ground})} interact in the triplet $s$-state and
the total spin of the pair 
is an integer, the pairs behave as bosons \cite{LIF80}. According to
the Bose-Einstein statistics two bosons attract each other and
the presence of the BCS pairing favors small interatomic spacing,
hence, {\it enhancing} the optical interactions and the 
light scattering.

The line shift induced by $\Sigma_2$ [Eq.~{(\ref{colshift})}]
is generated by the $s$-wave
interactions. As far as they can be considered
local on the scale of the optical wavelength in Eq.~{(\ref{h3})}
the collisions induce a local-field shift analogous 
to the Lorentz-Lorenz shift.

The optical line shift for the atomic sample is obtained from
Eq.~{(\ref{sus})}
\beq
{\cal S}/\gamma = 4\pi\bar\rho
+(\bar{u}_{g}-\bar{u}_{ge})\bar\rho_2(\up,\down)/\bar\rho
-6\pi{\rm Re}\( {\Sigma_1\over
\bar\alpha}\) \,,
\label{shift}
\eeq
where we have dropped the equal position coordinates in $\rho_2$,
used $\rho_2(\rv\nu,\rv\nu)=0$, and
defined the dimensionless variables $\bar\rho=
\rho/k^3$, $\bar\rho_2=\rho_2/k^6$, $\bar\delta=\delta/\gamma$, 
$\bar\alpha=-6\pi/(\bar\delta+i)$, and 
$\bar{u}_\xi=u_\xi k^3/\gamma$. The first two terms form
the local-field shift. For $^6$Li
the local-field shift due to the $s$-wave scattering in
Eq.~{(\ref{shift})} is larger
than the Lorentz-Lorenz shift, if $\gamma\lesssim 140
[1+(\Delta/\hbar u_g \rho)^2](a_g-a_{ge})/(a_0\lambda^3)$ $\mu m^3
s^{-1}$,
where $\lambda$ is the optical wavelength.
Because $(\Delta/\hbar u_g \rho)^2$ is expected to be of the 
order of one \cite{HOU97}, the local-field shift could 
strikingly depend on the BCS order parameter $\Delta$.

If the the effective range $r_u$ of the triplet $s$-wave 
potential in Eq.~{(\ref{h3})} is very short $r_u\ll 1/k$,
the resonant dipole-dipole interactions
may suppress the effect of the $s$-wave scattering on the line
shift just as they cancel the effect of the polarization 
self-energy \cite{RUO97c}. However, for a metastable state
$\gamma^{-1}$ may be large on the time-scale of the atomic 
interactions. In that case the collisional shift could be
observable even for very small $r_u$.

To calculate the nonlocal linewidth and line shift from
integral (\ref{c}) we need to evaluate the spatial 
correlation functions by using Eqs.~{(\ref{bogo})} and
{(\ref{bogo2})}. For instance, 
the anomalous expectation value reads
\beq
\<\psi_\down(\rv)\psi_\up(0)\>={1\over V}\sum_\kv e^{i\kv \cdot
\rv}{\Delta\over 2E_\kv}(1-\bar{n}_{\alpha \kv}-
\bar{n}_{\beta \kv})\,.
\label{pairsum}
\eeq
The chemical potential is solved from $\rho_\nu=\rho_\nu(\bar\mu)$.
Here $\<\psi_\down(0)\psi_\up(0)\>=-\Delta/(\hbar u_g)$ is 
ultraviolet-divergent resulting from
the assumption of the contact two-body interaction in
Eq.~{(\ref{ground})}. This interaction is momentum independent and
is not valid at high energies. To estimate the pairing
we remove the high-energy divergence by replacing $\hbar u_g$ by the
two-body T-matrix obtained from the Lippmann-Schwinger 
equation \cite{HOU97}. 
This is done by
subtracting from Eq.~{(\ref{pairsum})} $\sum \exp(i\kv\cdot\rv)
\Delta/(2\xi_\kv V)$. As argued in 
Ref. \cite{HOU97} the use of the T-matrix may seriously
underestimate the overlap in the case of large
scattering length $|a_g|\gg r_u$. Hence, we calculate 
Eq.~{(\ref{pairsum})} also by introducing a high-momentum cut-off~$k_c$.

We have plotted the line shift from Eq.~{(\ref{shift})} and
the linewidth $\Gamma/\gamma=1-6\pi {\rm Im}(\Sigma_1/\bar\alpha)$
by assuming, for simplicity, $\bar{u}_g=\bar{u}_{ge}$, $\lambda=
900$ nm, and a moderate value $a_g=-1200a_0$.
For the gap parameter at $T=0$ we use the weak coupling
approximation $\Delta\simeq 1.76 k_B T_c$ \cite{LIF80,HOU97}, where
\beq
k_BT_c\simeq {8\epsilon_F\over\pi} e^{\gamma-2}
\exp\[-{\pi\over2k_F|a|}\]\,,
\eeq
with $\gamma\simeq0.5772$ and $k_F=(6\pi^2\rho)^{1/3}$.

\begin{figure}
\begin{center}
\epsfig{file=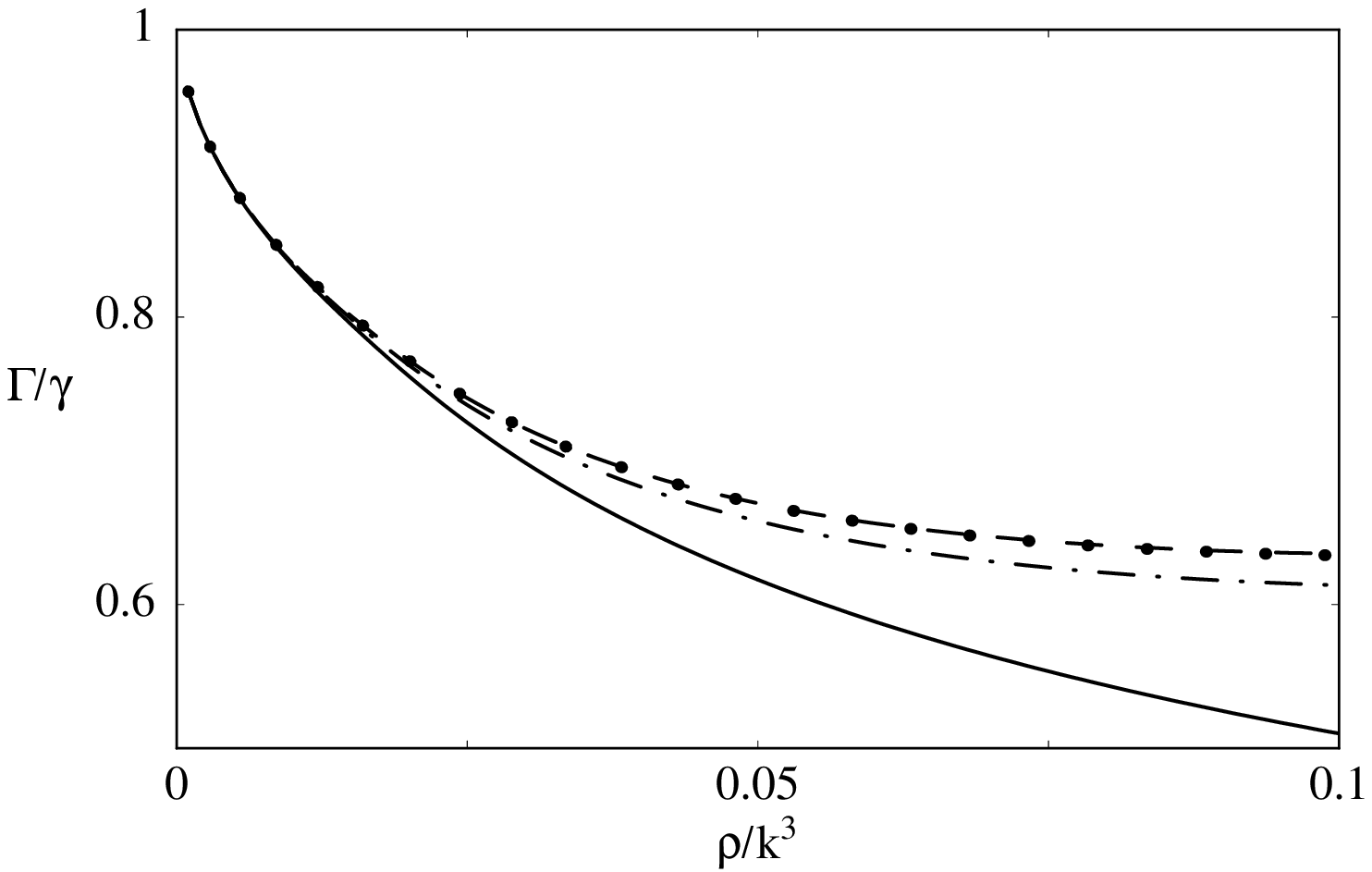,width=7.5cm}
\epsfig{file=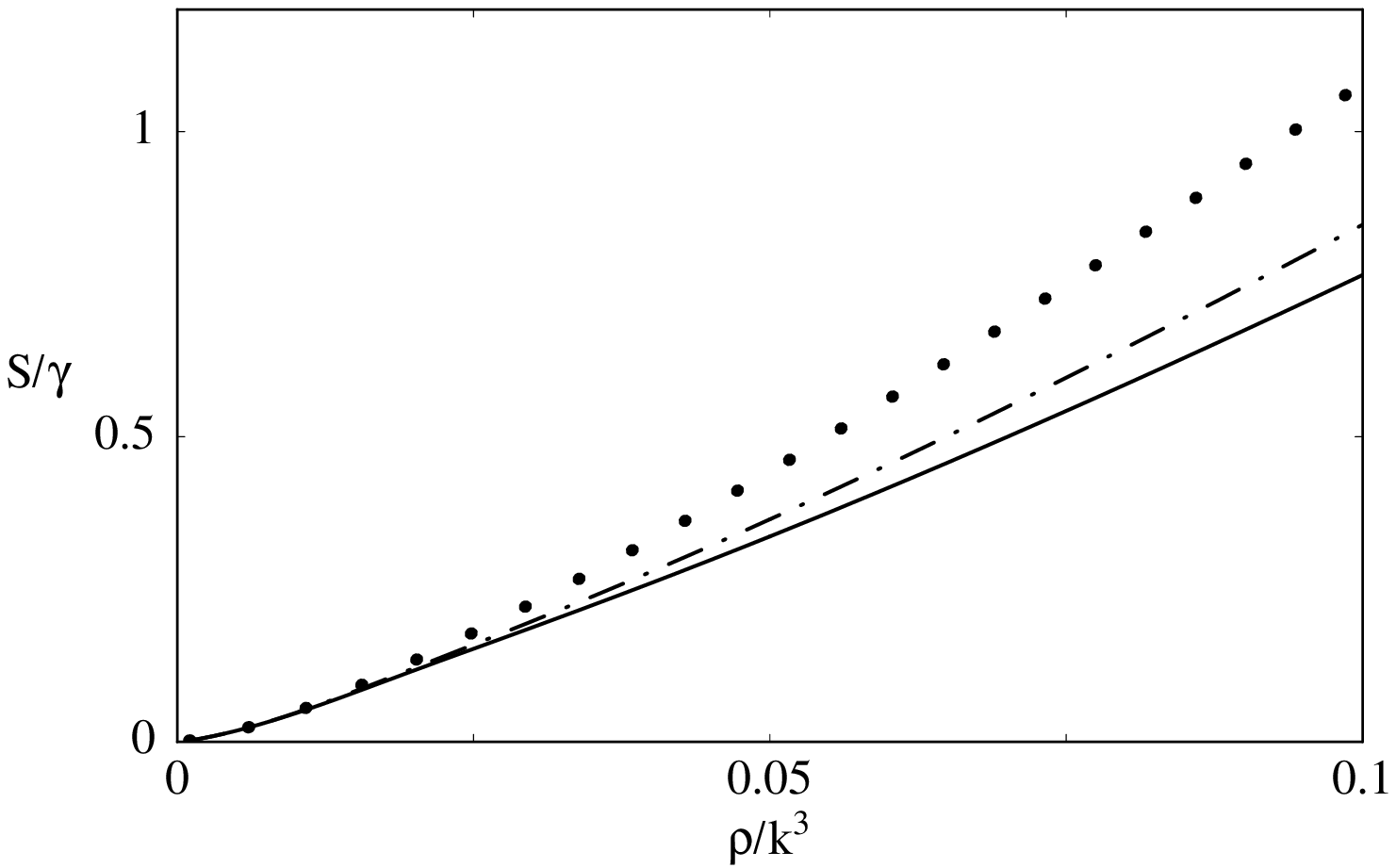,width=7.5cm}
\end{center}
\caption{The optical (a) linewidth and (b) 
line shift as a function of the atom density per cubic optical
wave number of the driving light. The dashed-dotted line corresponds
to the regularization by the two-body T-matrix, the dotted 
by the cut-off $k_c=0.01a_0^{-1}$, 
and the dashed line is the unregularized case. The solid line
has $\Delta=0$.
}
\label{fig:1}
\end{figure}

In Fig. \ref{fig:1} (a) the solid line represents the linewidth 
in the absence of the superfluid state ($\Delta=0$). The line
narrows as a function of the density \cite{RUO99c}. The presence
of the superfluid state broadens the line.
The linewidth is finite even without the regularization
in the anomalous correlation (the dashed line).
This is because the dipole radiation already involves a 
high-frequency cut-off \cite{RUO97a} that
regularizes small $r$ behaviour.
We have also plotted the case with the cut-off
$k_c=1/r_u$ (the dotted line)
with the realistic value $r_u=100a_0$ of the triplet $s$-wave
potential \cite{HOU97}. We found that the linewidth is 
almost independent of the cut-off from $r_u=500a_0$ to $r_u=0$. 
The line shift from the unregularized anomalous correlation 
in integral (\ref{c})
diverges logarithmically for small $r$. Although the radiation
kernel (\ref{eq:GDF}) involves a cut-off, the Lamb shift is not treated
rigorously. However, for the present purposes we may at least obtain
an estimate for the shift by using the T-matrix
or the cut-off $k_c=1/r_u$ as in the case of the linewidth. 
For the BCS state, even with $\bar{u}_g=\bar{u}_{ge}$ and moderately
small $|a_g|$, also the line shift is increased. 

We studied the interaction of light with a two-species 
atomic superfluid gas. The analysis of the quasi-particles followed
the standard BCS theory \cite{LIF80}.
We assumed a translationally invariant system.
Atoms in a harmonic trap
may be considered as locally homogeneous \cite{HOU97} provided that
the trap length scale $l=(\hbar/m\omega)^{1/2}$ is much larger than
the correlation length, which for intraspecies is $\xi_{\nu\nu}
\sim1/k_F$ and for the interspecies $\xi_{\up\down}\sim \epsilon_F/(
\Delta k_F)$ \cite{LIF80}. Other notable assumptions were zero
temperature and low atom density. In the future the present work 
could be extended to larger values of $|a_g|$ and $\bar\rho$
by going beyond the BCS 
weak coupling limit and by including
the cooperative optical linewidths and line shifts \cite{RUO97a}.

We acknowledge financial support from EU through the
TMR Network ERBFMRXCT96-0066.


\begin{references}
\bibitem{becexp} M. H. Anderson {\it et al.},  Science {\bf 269}, 198 (1995);
K. B. Davis {\it et al.}, Phys. Rev. Lett. {\bf  75}, 3969 (1995);
C. C. Bradley {\it et al.}, Phys. Rev. Lett. {\bf 75}, 1687 (1995).

\bibitem{JAV95b} J. Javanainen and J. Ruostekoski, Phys. Rev. A {\bf 52}, 
3033 (1995).

\bibitem{STO96} H. T. C. Stoof {\it et al.}, Phys. Rev. Lett. {\bf 76},
10 (1996).

\bibitem{HOU97} M. Houbiers {\it et al.}, Phys Rev. A {\bf 56}, 4864 (1997).

\bibitem{MAR98} B. DeMarco and D. S. Jin, Phys. Rev. A {\bf 58}, R4267
(1998); B. DeMarco {\it et al.}, e-print cond-mat/9812350.

\bibitem{BAR96} M.A. Baranov {\it et al.}, Zh. Eksp. Teor. Fiz. {\bf 64}, 273
(1996) [Sov. Phys. JETP Lett. {\bf 64}, 301 (1996)]; A.G.W. Modawi and
A.J. Leggett, J. Low. Temp. Phys. {\bf 109}, 625 (1998).

\bibitem{BRU98b} G. Bruun {\it et al.}, e-print cond-mat/9810013.

\bibitem{JAV99} J. Javanainen {\it et al.},
Phys. Rev. A {\bf 59}, 649 (1999).

\bibitem{RUO99c} J. Ruostekoski and J. Javanainen, e-print cond-mat/9902172.

\bibitem{LIF80} E.M. Lifshitz and L.P. Pitaevskii, 
{\it Statistical Physics},
Part II (Pergamon, Oxford, 1980).

\bibitem{FET71}  A.L. Fetter and J.D. Walecka, {\it Quantum
Theory of Many-Particle Systems} (McGraw-Hill, New York, 1971). 

\bibitem{COH89}
C. Cohen-Tannoudji, J. Dupont-Roc, and G. Grynberg, {\it Photons and Atoms
} (Wiley, New York, 1989); M. Lewenstein {\it et al.}, Phys. Rev. A {\bf
50}, 2207 (1994).

\bibitem{RUO97c} J. Ruostekoski and J. Javanainen, Phys. Rev. A {\bf 56},
2056 (1997).

\bibitem{RUO97a} J. Ruostekoski and J. Javanainen, Phys. Rev. A {\bf 55},
513 (1997).

\bibitem{JAC75} J.D. Jackson, {\it Classical Electrodynamics}, 2nd
ed. (Wiley, New York, 1975).

\bibitem{MOR95} O. Morice, Y. Castin, and J. Dalibard, Phys. Rev. A
{\bf 51}, 3896 (1995).

 
\end{references}
\end{document}